\documentclass[aps,prl, superscriptaddress,amsmath,amssymb,twocolumn,amsfonts,floatfix,longbibliography]{revtex4-2}
\usepackage{mathtools}
\usepackage{braket}
\usepackage[dvipsnames]{xcolor}
\usepackage{float}
\usepackage{subfigure}
\usepackage{upgreek}
\usepackage{pgffor}
\usepackage{silence}
\usepackage[colorlinks=true,linktoc=page,linkcolor=blue,citecolor=magenta,urlcolor=Fuchsia]{hyperref}
\usepackage[normalem]{ulem}
\usepackage{sidecap,tikz}
\usepackage{orcidlink}
\usepackage{makecell}
\usepackage{mathrsfs}
\usepackage{comment}

\newcommand{\beq}{\begin{equation}}
\newcommand{\eeq}{\end{equation}}
\newcommand{\bea}{\begin{eqnarray}}
\newcommand{\eea}{\end{eqnarray}}

\newcommand{\eqn}[1]{Eq.~(\ref{#1})}
\newcommand{\fig}[1]{Fig.~\ref{#1}}

\mathchardef\mhyphen="2D

\begin{document}

\title{Probing persistent spin textures through nonlinear magnetotransport}

\author{Neelanjan Chakraborti}
\thanks{These authors contributed equally}
\affiliation{Department of Physics, Indian Institute of Technology, Kanpur 208016, India}

\author{Akash Dey}
\thanks{These authors contributed equally}
\affiliation{Department of Physics, National Institute of Science Education and Research, Jatni, 752050, India}
\affiliation{Homi Bhabha National Institute, Training School Complex, Anushakti Nagar, Mumbai 400094, India}

\author{Snehasish Nandy}
\email{snehasish@phy.nits.ac.in}
\affiliation{Department of Physics, National Institute of Technology Silchar, Assam 788010, India}

\author{Sudeep Kumar Ghosh\,\orcidlink{0000-0002-3646-0629}}
\email{skghosh@iitk.ac.in}
\affiliation{Department of Physics, Indian Institute of Technology, Kanpur 208016, India}

\author{Kush Saha}
\email{kush.saha@niser.ac.in}
\affiliation{Max-Planck Institute for the Physics of Complex Systems, Noethnitzer Str. 38, 01187, Dresden, Germany}
\affiliation{Department of Physics, National Institute of Science Education and Research, Jatni, 752050, India}
\affiliation{Homi Bhabha National Institute, Training School Complex, Anushakti Nagar, Mumbai 400094, India}

\begin{abstract}
Persistent spin textures (PST) are special spin configurations in spin-orbit-coupled systems in which the spin polarization acquires a symmetry-enforced momentum-independent orientation, leading to exceptionally long spin lifetimes and persistent spin helices. Identifying direct experimental probes of PST, however, remains challenging because conventional quantum-geometric responses are strongly suppressed in this regime. Here, we show that PST systems isolate spin-rotation quantum geometry, which manifests through distinctive nonlinear magnetotransport responses. Using both a fine-tuned Rashba-Dresselhaus two-dimensional electron gas and a symmetry-enforced cubic spin-splitting model realizing PST, we demonstrate that PST suppresses conventional and Zeeman quantum-geometric contributions, leaving the spin-rotation quantum geometric tensor as the sole source of nonlinear magnetic-current and spin-magnetization responses. Remarkably, the nonvanishing response components exhibit identical direction-independent behavior as a function of chemical potential, providing a distinctive signature of PST. We further show that, in the Rashba-Dresselhaus two-dimensional electron gas at the PST point, these qualitative signatures remain robust even in the presence of a cubic Dresselhaus term that breaks the exact SU(2) symmetry. Our results establish nonlinear magnetotransport as an experimentally accessible probe of PST and their underlying spin-rotation quantum geometry.

\end{abstract}
\maketitle

Persistent spin textures (PST) constitute a distinctive regime of spin-orbit-coupled systems in which spin dynamics are governed by a conserved spin projection~\cite{SchliemannPST,Daniel}. Initially identified in two-dimensional electron gas (2DEG) systems with equal Rashba and Dresselhaus spin-orbit couplings, the PST condition stabilizes a persistent spin helix and dramatically enhances spin lifetimes~\cite{Bernevig_2006}. Beyond this fine-tuned setting, PST has also been shown to emerge intrinsically through crystalline-symmetry constraints~\cite{Tao2018,Zhao2020,kilic2025universal}. The momentum-independent spin polarization of Bloch states in PST systems suppresses conventional quantum-geometric responses~\cite{yu2025quantum,jiang2025revealing,verma2026quantum}, making direct experimental probes of PST challenging.

Quantum geometry, encoded in the quantum geometric tensor (QGT), provides a unified framework for characterizing Bloch states~\cite{yu2025quantum,jiang2025revealing,verma2026quantum}. Its imaginary part defines the Berry curvature, while its real part defines the quantum metric. These geometric quantities play a central role in anomalous transport, nonlinear responses, and geometric contributions to the superfluid density in flat-band superconductors~\cite{yu2025quantum,jiang2025revealing,verma2026quantum,Mitscherling_2020,Postlewaite2024,Nandy_2022,Nandy_2023,ulrich2025quantum,Akash2025,sayan2025,Antebi2024,Mitscherling2022,Peotta2015,Liang2017,Iguchi_2024,Hattori_2025,Nandy_2025}. In its conventional form, however, the QGT captures only momentum-space variations of Bloch wavefunctions. Consequently, the momentum-independent spin structure of PST~\cite{SchliemannPST,Daniel} strongly suppresses conventional quantum-geometric responses, suggesting that transport probes rooted in conventional quantum geometry may become ineffective in the PST regime. This naturally raises a central question: can any quantum-geometric transport response survive in PST systems and thereby provide a direct probe of PST?

\begin{figure}[!t]
\centering
\includegraphics[width=0.95\columnwidth]{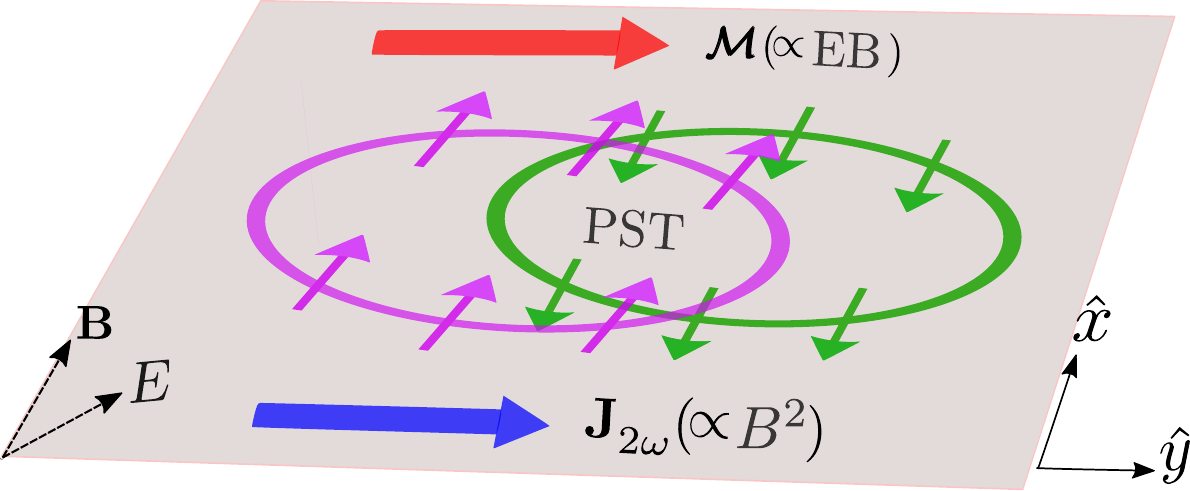}
\caption{\textbf{Schematic of nonlinear magnetic current and spin-magnetization responses in the PST regime:} A system in the PST regime, driven by electric ($\mathbf{E}$) and magnetic ($\mathbf{B}$) fields, exhibits a nonlinear spin-magnetization response of order $EB$ and a nonlinear magnetic current response of order $B^2$. Both responses originate from a finite spin-rotation quantum geometry, while all other quantum-geometry-driven responses vanish.}
\label{fig:schematic}
\end{figure}

Here, we answer this question in the affirmative by showing that PST systems can indeed be probed through nonlinear magnetotransport governed by generalized quantum geometry. Recent extensions of quantum geometry to spin-orbit-coupled systems have generalized the QGT to include spin-space rotations, leading to two additional geometric tensors: the spin-rotation QGT (SRQGT) and the Zeeman QGT~\cite{Xiang_2025_PRL,Chakraborti2025,chakraborti2026magnetization,xiang_2025_em}. While the SRQGT characterizes purely spin-space rotations, the Zeeman QGT captures the coupled effect of spin rotations and momentum-space translations. We show that PST systems suppress both the conventional and Zeeman quantum-geometric contributions while retaining a finite spin-rotation quantum geometry. Consequently, nonlinear magnetotransport becomes governed entirely by the SRQGT, thereby providing a direct probe of PST. Using both the equal Rashba-Dresselhaus 2DEG and a symmetry-enforced PST phase generated by cubic spin splitting, we demonstrate distinctive nonlinear magnetic-current and spin-magnetization signatures of PST. Remarkably, the nonzero response components exhibit identical direction-independent behavior as a function of chemical potential, providing a characteristic transport signature of PST. We further show that, for the 2DEG at the PST point, these qualitative signatures remain robust even in the presence of a cubic Dresselhaus perturbation that breaks the exact SU(2) symmetry.\\

\begin{figure}[!b]
\centering
\includegraphics[width=0.99\columnwidth]{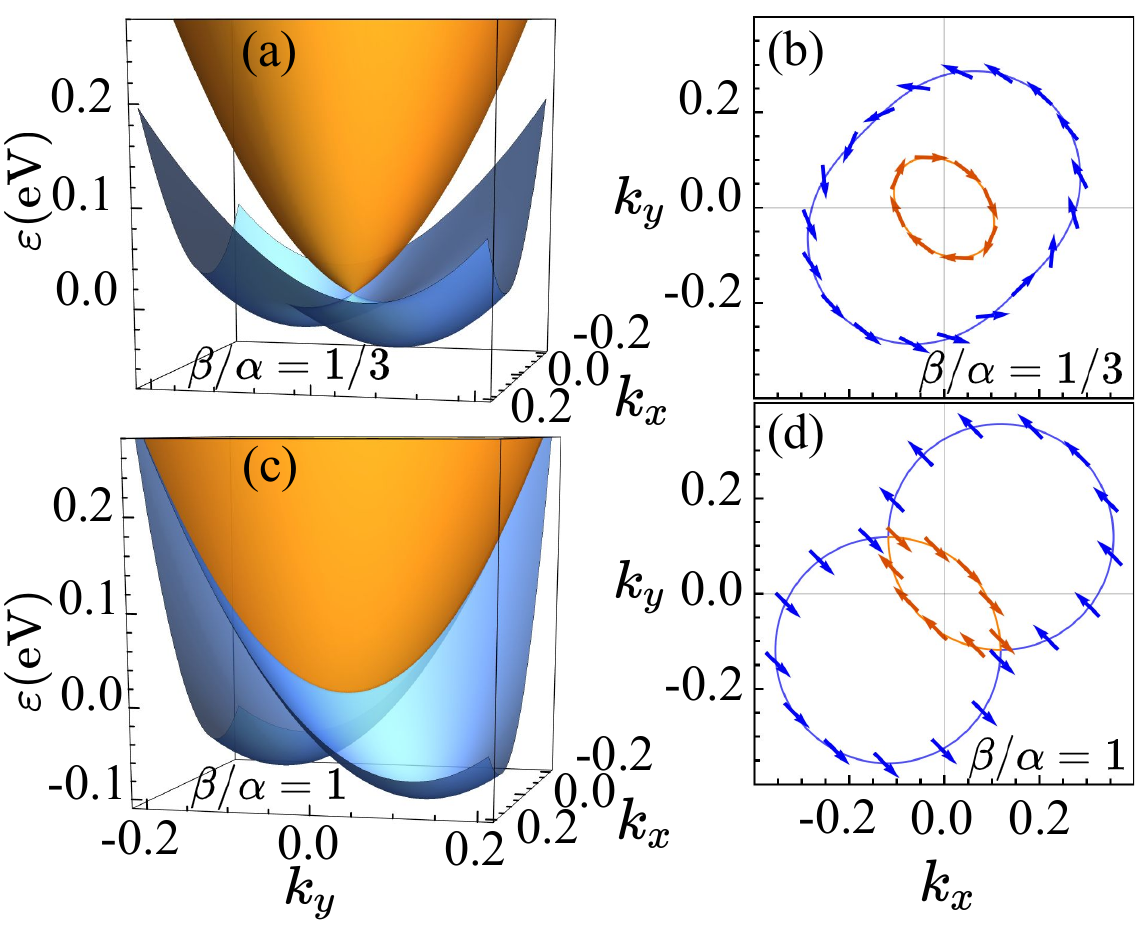}
\caption{\textbf{Energy dispersions and spin textures on Fermi surfaces of a 2DEG:}
(a, c) Energy dispersions for $\beta/\alpha=1/3$ and $\beta/\alpha=1$, respectively. The isolated band-touching point at zero energy for $\alpha \neq \beta$ transforms into a nodal line for $\alpha=\beta$. (b, d) Spin textures on Fermi contours with $\mu = 0.03~\mathrm{eV}$ for $\beta/\alpha=1/3$ and $\beta/\alpha=1$, respectively. Here, we have chosen $\alpha = 0.6~\mathrm{eV\cdot\mathring{A}}$ and $m=0.14~\mathrm{eV^{-1}\cdot\mathring{A}^{-2}}$. 
}\label{fig:2DEG_disp}
\end{figure}

\noindent\textit{Spin-rotation quantum geometry:} While quantum geometry traditionally characterizes electronic wavefunctions through momentum-space variations, a complete description of Bloch states $|u^{\zeta}_{m\mathbf{k}}\rangle$ in spin-orbit coupled systems must also incorporate spin-space rotations~\cite{Xiang_2025_PRL}. This unified framework generates three distinct QGTs by combining the position ($r$) and Pauli spin ($\sigma$) operators: the conventional QGT ${\mathscr G}_{mp}^{ab} = r_{mp}^{a} r_{pm}^{b}$~\cite{Xiao_2010,Jiang_2025,Gao_2014,Du_2021}, the Zeeman QGT ${\mathscr Z}_{mp}^{ab} = r_{mp}^{a} \sigma_{pm}^{b}$~\cite{Xiang_2025_PRL,chakraborti2025_arxiv}, and the SRQGT ${\mathscr S}_{mp}^{ab} = \sigma_{mp}^{a} \sigma_{pm}^{b}$~\cite{Xiang_2025_PRL,chakraborti2025_arxiv}. Here, $a$ and $b$ represent spatial indices, while $m$ and $p$ denote the Bloch band indices. The real and imaginary components of these tensors define their respective quantum metrics and Berry curvatures. In particular, the spin-rotation quantum metric (SRQM) $\mathscr{R}^{ab}_{mp}$ and the spin-rotation Berry curvature (SRBC) $\Lambda^{ab}_{mp}$~\cite{Jia_2025} are pivotal in governing the unique responses of spin-orbit coupled systems, providing a robust probe for PST as described in the following. Under time-reversal ($\mathcal{T}$) and inversion ($\mathcal{P}$) symmetries, the SRQM is even, whereas the SRBC is odd in both.\\


\noindent \textit{PST in a two-dimensional electron gas (2DEG):} 
In noncentrosymmetric semiconductor quantum wells and heterostructures, including GaAs/AlGaAs and InAs-based 2DEGs~\cite{Winkler2003,Ganichev2004,Shojaei2016,Walser2012PSH,Koralek2009PSH}, Rashba and Dresselhaus spin-orbit interactions~\cite{Bychkov1984,Dresselhaus1955} naturally arise from structural and bulk inversion asymmetry, respectively. The resulting low-energy Hamiltonian describing the 2DEG is given by (we set $e=\hbar=1$),
\begin{align}
H = \frac{k^2}{2m} + \alpha \left(k_y \sigma_x - k_x \sigma_y\right) + \beta \left(k_x \sigma_x - k_y \sigma_y\right), \label{eq:Hamiltonian}
 \end{align} 
where $\sigma_i$ are the spin Pauli matrices, $\alpha$ and $\beta$ are the strengths of Rashba and Dresselhaus spin–orbit coupling respectively. The system preserves time-reversal symmetry while breaking inversion symmetry. The dispersions and the spin textures on the Fermi surfaces of the model are shown in \fig{fig:2DEG_disp}. A particularly intriguing regime arises at $\alpha=\beta$, where an emergent SU(2) symmetry stabilizes PST~\cite{Schliemann2003,Bernevig_2006}. In this limit, the effective spin–orbit field reduces to $\mathbf{\mathcal{B}}_{\mathrm{eff}}=\alpha (k_x+k_y)(1,-1)$, whose direction becomes momentum independent, resulting in unidirectional PST, as illustrated in Fig.~\ref{fig:2DEG_disp}(d)~\cite{Tao2018,Dey2025}. The emergent SU(2) symmetry implies the existence of a conserved spin operator $\Sigma=(\sigma_x-\sigma_y)/\sqrt{2}$, which renders the spinor component of the Bloch eigenstates independent of the crystal momentum $\mathbf{k}$. This becomes transparent in the rotated spin basis defined by $\Sigma$, where the spin–orbit coupling takes the form $H_{\mathrm{SOC}}(\mathbf{k})=\sqrt{2}\alpha (k_x+k_y)\Sigma$, satisfying $[H,\Sigma]=0$.  

A notable feature of the PST regime is that both the conventional and Zeeman quantum geometric tensors vanish identically (see Supplemental Material (SM) for details). This is a direct consequence of the momentum-independent spinor structure of the Bloch eigenstates, since these geometric quantities originate from the momentum-space variation of the Bloch wave functions. As a result, all the canonical transport responses driven by these quantum geometry vanish in the case of PST. In striking contrast, the SRQGT remains finite in this regime. In particular, the in-plane spin–orbit coupling field constrains the band-resolved SRBC to lie strictly out of plane ($\Lambda^{xz}_{\pm \mp},\Lambda^{yz}_{\pm \mp} \neq 0$) and simultaneously forces the out-of-plane component of the band-resolved SRQM ($\mathscr{R}^{xz}_{\pm \mp},\mathscr{R}^{yz}_{\pm \mp} = 0$) to vanish. The magnitudes of both the SRQM and SRBC are momentum independent; however, unlike the SRQM, which retains a fixed sign, the sign of the SRBC depends on momentum: 
\begin{equation}
\mathscr{R}^{ab}_{\pm\mp} = 1/2 \;\;\&\;\; \Lambda_{\pm\mp}^{az}=\mp \sqrt{2} \;{\rm sgn}(k_x+k_y) .
\label{eq: SRBC components}
\end{equation}
This establishes the SRQGT as the sole surviving quantum geometric quantity at the PST point-- one of the central results of the paper.

Moving beyond the PST regime, for generic couplings ($\alpha \ne \beta$), the effective spin-orbit field $\mathbf{\mathcal{B}}_{\mathrm{eff}} = [\beta k_x +\alpha k_y, -(\alpha k_x + \beta k_y)]$ is momentum-dependent, producing anisotropic spin textures on the Fermi contours as depicted in Fig.~\ref{fig:2DEG_disp}(b). In this regime, the strictly in-plane nature of $\mathbf{\mathcal{B}}_{\mathrm{eff}}$ causes all components of conventional Berry curvature $\Omega^{ab}_{\pm\mp}$ and Zeeman quantum metric (ZQM) $\mathcal{F}^{ab}_{\pm\mp}$, as well as $\Lambda^{xy}_{\pm\mp}$, $\mathscr{R}^{xz}_{\pm\mp}$, and $\mathscr{R}^{yz}_{\pm\mp}$, to vanish, while the remaining components of the different QGTs remain finite (see SM for details). \\

\begin{figure}
\centering
\includegraphics[width=\columnwidth]{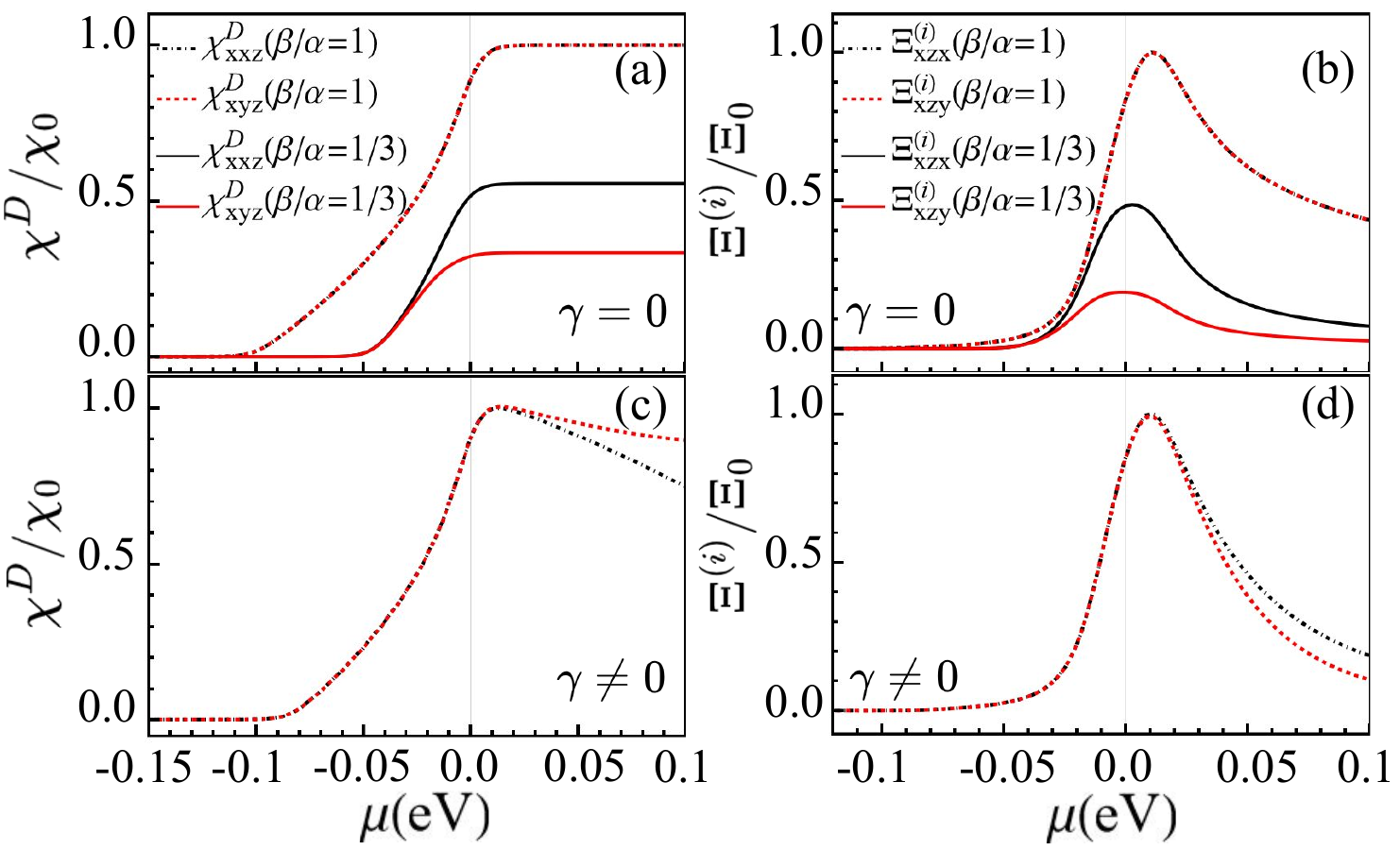}
\caption{\textbf{Nonlinear magnetic current and spin-magnetization responses of PST in a 2DEG:}
(a) Behavior of the nonzero displacement nonlinear magnetic  conductivity components, $\chi^{D}_{xxz}$ and $\chi^{D}_{xyz}$, with $\mu$ at the PST point $\beta/\alpha=1$ (dashed lines) and for $\beta/\alpha=1/3$ (solid lines) with no cubic Dresselhaus term, i.e. $\gamma = 0$. (b) Corresponding variation of the non-vanishing components of the nonlinear spin-magnetization response tensor, $\Xi^{(i)}_{xzx}$ and $\Xi^{(i)}_{xzy}$. (c,d) Same quantities with a finite cubic Dresselhaus term $\gamma = 10~\mathrm{eV\cdot\mathring{A}^3}$ at the PST point. We note that the coalescence of the different response components is lifted only after certain positive values of $\mu$. Here $\chi_0 = \left(\frac{g\mu_B}{2}\right)^2|\chi^{D}_{\text{max}}| \,\mathrm{A^{-1}\,m\,T^{2}}$ and $\Xi_0=\Xi^{(i)}_{max}~ \mathrm{{nm}^{-1}V^{-1}T^{-1}}$. We have chosen $\alpha=0.6~\mathrm{eV\cdot\mathring{A}}$, $T = 50~\mathrm{K}$ and $\omega = 10^{13}~\mathrm{Hz}$.
}\label{fig:2DEG_response}
\end{figure}

\noindent \textit{Magnetotransport responses of PST:} 
To demonstrate transport signatures of a finite SRQGT as a probe of the PST, we investigate the magnetotransport response of a 2DEG driven by time-dependent electric and magnetic fields. In the PST regime, the linear magnetoelectric responses~\cite{Xiao2010,Xiang_2025_PRL,Goldberg_1954,Ghorai_2025,Nakayama_2013,Cortijo_2016,Monteiro_2015} governed by the conventional and Zeeman QGTs vanish. The absence of such linear responses therefore necessitates the consideration of nonlinear transport responses governed by generalized quantum geometry. We consequently focus on the second-order magnetic charge current $(\sim B^2)$~\cite{chakraborti2026} and the nonlinear spin-magnetization response bilinear in electric and magnetic fields $(\sim EB)$~\cite{Jia_2025}, which are governed by the SRQGT and remain finite at the PST point. A time-dependent magnetic field with frequency $\omega$ induces a nonlinear magnetic current~\cite{chakraborti2026}: 
\beq
J^{a}_{2\omega} = \chi_{abc}^{M} B^b(t) B^c(t),
\eeq
where the conductivity tensor $\chi_{abc}^{M}$ decomposes as $\chi_{abc}^{M} = \chi_{abc}^{C} + \chi_{abc}^{D}$. The conduction contribution $\chi_{abc}^{C}$ is governed by the SRQM, while the displacement contribution $\chi_{abc}^{D}$ is governed by the SRBC (see SM for details). These are given by~\cite{chakraborti2026},
\bea
\chi_{abc}^{C} &= \left(\frac{g\mu_B}{2}\right)^2 \sum_{m\neq p} f_{m} \frac{\partial \varepsilon_{mp}}{\partial k^a} \left(\frac{\mathscr{R}_{mp}^{bc}}{\varepsilon_{mp}^2}\right), \label{eq:conductivitiesa}\\ 
\chi_{abc}^{D} &= \left(\frac{g\mu_B}{2}\right)^2 \sum_{m\neq p} f_{m} \frac{\partial \varepsilon_{mp}}{\partial k^a} \left(\frac{\Lambda_{mp}^{bc}}{\omega \varepsilon_{mp}}\right),\label{eq:conductivitiesb} 
\eea
where $f_m$ is the Fermi–Dirac distribution and $\varepsilon_{mp}=(\varepsilon_m-\varepsilon_p)$ is the interband energy difference.

We calculate the nonlinear spin magnetization $(\pmb{\mathcal{M}})$ within the framework of quantum kinetic theory~\cite{kamal_2022_prl}. Our results are consistent with previous studies based on the Keldysh Green’s function formalism~\cite{Jia_2025}. It can be written as
\beq
\mathcal{M}^a = \mu_B \Xi_{acb} E^b B^c,
\eeq
where the magnetization response tensor $\Xi_{acb}$ decomposes as $\Xi_{acb} = \left[\Xi^{(i)}_{acb} + \Xi^{(e)}_{acb}\right]$. The intrinsic and extrinsic response tensors are given by~\cite{Jia_2025}
\bea
\hspace{-0.3cm}\Xi^{(i)}_{acb} &=& g \mu_B \sum_{p m} \int_{\mathbf{k}} \frac{2 f_p}{\varepsilon_{mp}^2}
\left[\Delta\sigma_{pm}^{c}\,\mathcal{F}_{mp}^{ba} + \frac{\partial \varepsilon_{mp}}{\partial k^b} \frac{\Lambda_{mp}^{ac}}{2\varepsilon_{mp}}
\right] \label{eq:mag_int}\\
\hspace{-0.3cm}\Xi^{(e)}_{acb} &=& - g \mu_B \sum_{p \ne m} \int_{\mathbf{k}}
f_p \,\nabla_{b} \left(\frac{2 \mathscr{R}_{pm}^{ac}}{\varepsilon_{pm}} \right)\label{eq:mag_ext}.
\eea
The extrinsic contribution $\Xi^{(e)}_{acb}$ is governed solely by the SRQM and is purely a Fermi-surface quantity. In contrast, the intrinsic contribution $\Xi^{(i)}_{acb}$ involves the SRBC and ZQM, and is a Fermi-sea quantity.

Figure \ref{fig:2DEG_response} illustrates the nonlinear responses of the 2DEG (\eqn{eq:Hamiltonian}) as a function of the chemical potential. Owing to time-reversal symmetry, the conduction magnetic conductivity $\chi^{C}$ vanishes identically. The momentum derivative of the interband energy difference in Eq.~(\ref{eq:conductivitiesa}) is odd under $\mathbf{k}\rightarrow -\mathbf{k}$, making the integrand odd in momentum, while the Fermi function remains even. Consequently, the Brillouin-zone integral vanishes, with each band contribution vanishing separately. In contrast, the integrand of $\chi^{D}$ is even in $k$ as $\Lambda(-{\bf k})=-\Lambda({\bf k})$, hence yields a finite contribution. This conclusion holds for both at the PST ($\alpha=\beta$) and non-PST point ($\alpha\ne\beta$) as evident from Fig.~\ref{fig:2DEG_response}a. Notably, in the PST regime all nonvanishing $\chi^D$ components become identical, in sharp contrast to the non-PST case, owing to the momentum-independent nature of the SRBC at the PST point [see Eq.~(\ref{eq: SRBC components})]. This serves as a {\it smoking gun} signature of PST in a spin-orbit coupled system. 

The behavior of $\chi^{D}$ now can be understood from the band structure shown in Fig.~\ref{fig:2DEG_disp}(a) and (c). For $\mu < 0$, the response is dominated by a single outer band, leading to a monotonic increase in magnitude. For $\mu > 0$, the inner and outer bands contribute with opposite signs, resulting in a partial cancellation. As a consequence, $\chi^{D}$ approaches a saturated value at larger $\mu$, since the competing contributions grow at comparable rates governed by $(f_m - f_n)$. Moreover, the PST regime exhibits an overall enhancement of $\chi^{D}$ relative to the non-PST case, reflecting the larger SRBC contributions (see Fig.~\ref{fig:2DEG_response}a).

A similar analysis applies to the nonlinear spin magnetization response tensor $\Xi_{acb}$. The extrinsic contribution $\Xi^{(e)}_{acb}$ vanishes under TRS, as its integrand in \eqn{eq:mag_ext} is odd in momentum and cancels over the Brillouin zone, analogous to $\chi^{C}$. In contrast, the intrinsic contribution $\Xi^{(i)}_{acb}$ remains finite because its integrand in \eqn{eq:mag_int} is even in momentum (see \fig{fig:2DEG_response}b). Similar to $\chi^{D}$, the PST regime ($\alpha=\beta$) yields identical tensor components for $\Xi^{(i)}$ , while the non-PST case retains anisotropy, providing a clear signature of PST.

Figure~\ref{fig:2DEG_disp}(a) and (c) further clarifies the behavior of $\Xi^{(i)}_{acb}$. For $\mu < 0$, the response is dominated by the outer band and increases monotonically. For $\mu > 0$, in contrast to $\chi^{D}$, the magnitude of $\Xi^{(i)}_{acb}$ decreases with increasing $\mu$. This suppression arises from the additional weighting factor $(f_m - f_p)/\varepsilon_{mp}^2$, which reduces the total contribution of two bands as the $\mu$ increases. As before, the PST regime yields a larger magnitude of $\Xi^{(i)}_{acb}$ compared to the non-PST case, highlighting key role of the SRBC.

We further examine the robustness of our results in more realistic situations where the PST condition in the 2DEG is weakly perturbed. As a representative example, we include the cubic Dresselhaus term $H'=-\gamma (k_x^2 k_y\sigma_x - k_x k_y^2\sigma_y)$~\cite{Koralek2009}, which explicitly breaks the underlying SU(2) symmetry. The effective Hamiltonian then becomes $H_{\mathrm{eff}}=H+H'$, such that the exact PST condition is no longer satisfied. Nevertheless, the qualitative features of both the nonlinear magnetic-conductivity and spin-magnetization response tensors remain largely intact, as shown in Fig.~\ref{fig:2DEG_response}(c) and (d) respectively. Even for a relatively large value $\gamma = 10~\mathrm{eV\cdot\mathring{A}^3}$, the different tensor components exhibit only a small splitting at larger chemical potentials. Importantly, the characteristic coalescence of the response components persists in a regime of $\mu$ as long as the nodal structure in the dispersions remains present [see Fig.~\ref{fig:2DEG_response}(c) and (d)]. This robustness demonstrates that the nonlinear responses are governed predominantly by the underlying spin-rotation quantum geometry. Consequently, these nonlinear magnetotransport signatures provide a reliable diagnostic of PST in spin-orbit-coupled systems even in the presence of weak symmetry-breaking perturbations.

We note that a finite conduction response $\chi_{abc}^{C}$ governed by the SRQM can be induced without destroying the underlying PST by introducing a tilt potential along the $k_x$ direction for example. This perturbation breaks particle-hole and time-reversal symmetries while preserving the momentum-independent spin structure characteristic of PST. A distinctive signature emerges at the PST point $\alpha=\beta$: owing to the constant and finite SRQM, the nonlinear magnetic conductivities become direction independent, leading to a characteristic coalescence of all nonvanishing components of the magnetic-current response (see SM for details).
\\

\begin{figure}[!b]
\centering
\includegraphics[width=0.5\textwidth]{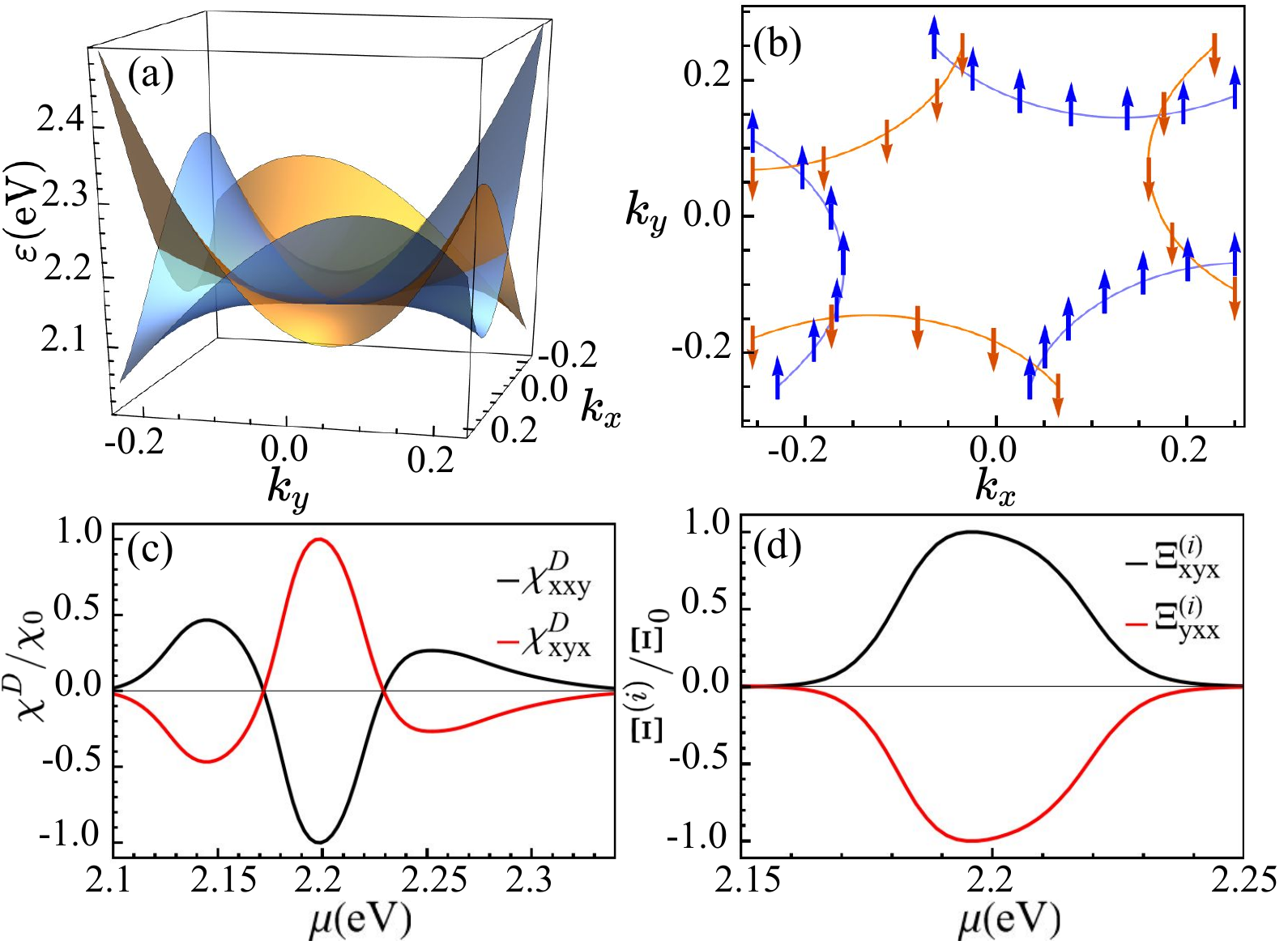}
\caption{\textbf{Nonlinear responses of a symmetry-enforced PST phase with cubic spin splitting:} (a) Energy dispersion and (b) Spin polarizations on the Fermi contours with $\mu=2.2$ eV showing PST in the system with cubic spin splittings (\eqn{eq:cubic_spin}). 
(c) Two representative nonzero components of the displacement nonlinear magnetic conductivity, $\chi_{xxy}^D$ and $\chi_{xyx}^D$. 
(d) Corresponding nonzero components of the nonlinear spin magnetization response tensor, $\Xi^{(i)}_{xyx}$ and $\Xi^{(i)}_{yxx}$ with $\Xi_0=|\Xi^{(i)}_{\text{max}}|~ \mathrm{{nm}^{-1}V^{-1}T^{-1}}$. The values of the parameters, chosen for a particular realization of \eqn{eq:cubic_spin} in Ge$_3$Pb$_5$O$_{11}$~\cite{Zhao2020}, are $\varepsilon_0 = 2.13~\mathrm{eV}$, $\Delta = 1.27~\mathrm{eV\!\cdot\!\mathring{A}^2}$, $\zeta = -5.24~\mathrm{eV\!\cdot\!\mathring{A}^3}$, $\lambda = -3.43~\mathrm{eV\!\cdot\!\mathring{A}^3}$. Here, $T = 50~\mathrm{K}$, and $\omega = 10^{13}~\mathrm{Hz}$.}
\label{fig: cubic_PST}
\end{figure}

\noindent \textit{PST in a system with purely cubic spin-splitting:}
To demonstrate that our results are not restricted to the Rashba-Dresselhaus 2DEG, we extend our analysis to a symmetry-enforced PST phase arising from purely cubic spin splitting in materials belonging to the $\bar{6}$ point group~\cite{Zhao2020}. The corresponding effective Hamiltonian~\cite{Zhao2020} is
\begin{equation}
H(\mathbf{k}) = \varepsilon_{0}+\Delta (k_x^{2}+k_y^{2})
+\Big[\zeta k_x (k_x^{2}-3k_y^{2})+\lambda k_y (3k_x^{2}-k_y^{2})\Big]\sigma_z ,
\label{eq:cubic_spin}
\end{equation}
with energy dispersion
$\varepsilon_\pm(\mathbf{k}) = \varepsilon_0 + \Delta (k_x^2 + k_y^2)
\pm [\zeta k_x (k_x^2 - 3k_y^2) + \lambda k_y (3k_x^2 - k_y^2)]$. Here, $\Delta$ characterizes the effective mass, while $\zeta$ and $\lambda$ denote the strengths of the cubic spin-splitting terms. The constant $\varepsilon_0$ can be used to fit the band structure of candidate materials realizing symmetry-enforced PST in the $\bar{6}$ point group~\cite{Zhao2020}.

Because the spin-dependent term in \eqn{eq:cubic_spin} is strictly proportional to $\sigma_z$, the spin orientation is locked along the out-of-plane direction throughout the Brillouin zone, realizing a PST that is independent of the specific values of $\zeta$ and $\lambda$ (see Fig.~\ref{fig: cubic_PST}(b)). As in the 2DEG case at the PST point, the eigenstates are momentum-independent, leading to the complete suppression of conventional and Zeeman QGTs. The SRQGT again emerges as the only finite quantum geometric quantity. In particular, the SRBC and SRQM remain finite and momentum-independent: $\mathscr{R}^{aa}_{\pm\mp} = 1$ \& $\Lambda^{ab}_{\pm\mp} = \pm 2$, where $a$ and $b$ denote in-plane spatial indices. Notably, despite being formally odd under time-reversal, the SRBC effectively behaves as a pseudo time-reversal-even quantity due to its momentum independence.

Since $\varepsilon_{+}(\mathbf{k}) = \varepsilon_{-}(-\mathbf{k})$, the Fermi-function difference is odd under $\mathbf{k}\rightarrow -\mathbf{k}$. Together with the momentum-independent SRQM, this renders the integrand of the conduction response [Eq.~(\ref{eq:conductivitiesa})] odd in momentum, leading to a vanishing Brillouin-zone integral. In contrast, time-reversal symmetry permits a finite displacement magnetic conductivity ($\chi^D$) governed entirely by the SRBC. The resulting displacement conductivities, such as $\chi^{D}_{xxy}$ and $\chi^{D}_{xyx}$ shown in \fig{fig: cubic_PST}(c), reflect the simple underlying spin-rotation geometry and exhibit a pronounced enhancement as the chemical potential approaches the {\it line-degeneracy} region near $\mu = 2.2~\mathrm{eV}$ [see \fig{fig: cubic_PST}(a)]. Note that $\chi^{D}_{xxy} = -\chi^{D}_{xyx}$ due to the antisymmetric structure of the SRBC, $\Lambda_{\pm \mp}^{xy} = - \Lambda_{\pm \mp}^{yx}$.

A similar structure emerges in the nonlinear spin-magnetization response. While the extrinsic contribution $\Xi^{(e)}_{acb}$ vanishes due to cancellation between the two bands, analogous to $\chi^{C}$, time-reversal symmetry permits a finite intrinsic contribution $\Xi^{(i)}_{acb}$ governed solely by the SRBC. Consequently, the intrinsic response components, such as $\Xi^{(i)}_{yxx}$ and $\Xi^{(i)}_{xyx}$ shown in Fig.~\ref{fig: cubic_PST}(d), display a strong enhancement near the line-degeneracy region, consistent with the behavior of $\chi^{D}$.

These results for the cubic spin-splitting model demonstrate that the nonlinear transport signatures governed by spin-rotation quantum geometry are not restricted to the Rashba-Dresselhaus 2DEG, but apply more generally to symmetry-enforced PST phases in noncentrosymmetric crystals~\cite{Zhao2020}. Recent works have shown that crystalline symmetries, particularly nonsymmorphic symmetries, can intrinsically stabilize momentum-independent spin polarization around high-symmetry points, lines, and planes in the Brillouin zone~\cite{Tao2018,kilic2025universal}. Our results therefore suggest that the predicted nonlinear magnetic-current and spin-magnetization responses should also arise in candidate PST materials such as BiInO$_3$, Be$_5$Pt, and OsSi~\cite{Tao2018,kilic2025universal}. Moreover, the recent discovery of crystalline-symmetry-protected PST phases in magnetic systems, particularly altermagnets~\cite{Tenzin2025,Campos2026PersistentAltermagnetism}, broadens the experimental landscape for observing these signatures.

Experimentally, PST and the associated persistent spin helix have primarily been probed through optical spin-grating techniques~\cite{Walser2012PSH,Koralek2009PSH}. In contrast, our work demonstrates that nonlinear magnetotransport provides a direct and experimentally accessible probe of PST through the underlying spin-rotation quantum geometry. This also offers a complementary route for accessing generalized quantum geometry in solid-state systems, where direct measurements of quantum geometry remain challenging despite recent experimental advances~\cite{kim2025direct,kang2025measurements}.

In summary, we have shown that PST systems provide a unique setting in which conventional and Zeeman quantum-geometric contributions are strongly suppressed, leaving nonlinear magnetotransport governed entirely by spin-rotation quantum geometry. Using both the Rashba-Dresselhaus 2DEG and symmetry-enforced PST phases generated by cubic spin splitting, we identified distinctive nonlinear magnetic-current and spin-magnetization signatures that remain robust even beyond the ideal SU(2)-symmetric limit. Our results establish a direct connection between PST and generalized quantum geometry, opening a pathway toward experimentally exploring spin-rotation quantum geometry in spin-orbit-coupled materials.

\textit{Acknowledgments:} We thank J. Mitscherling and Tarun K. Ghosh for useful discussions. N.~C. and A.~D. acknowledge the Council of Scientific and Industrial Research (CSIR), Government of India, for providing the SRF fellowship. S.~K.~G. and S.~N. acknowledge financial support from Anusandhan National Research Foundation (ANRF) erstwhile Science and Engineering Research Board (SERB), Govt. of India respectively via the Startup Research Grant: SRG/2023/000934 and the Prime Minister's Early Career Research Grant: ANRF/ECRG/2024/005947/PMS. K.~S acknowledge financial support from the Department of Atomic Energy (DAE), Govt. of India, through the project Basic Research in Physical and Multidisciplinary Sciences via RIN4001. 

\bibliography{PST}
\end{document}